\documentclass[11pt]{article}
\usepackage[dvips]{epsfig}

 \advance \textwidth by -2in
 \advance \textheight by -3in
\def\##1{{\bf{#1}}}
\def\=#1{\underline{\underline{#1}}}

\def\+#1{\underline{\bf #1}}
\def\*#1{\underline{\underline{\bf #1}}}

\def\.{\mbox{ \tiny{$^\bullet$} }}

\def\ux{\#{\hat u}_x}
\def\uy{\#{\hat u}_y}
\def\uz{\#{\hat u}_z}

\begin{document}

\begin{center}

 {\Large {\bf Total Internal Reflection of Evanescent Plane Waves}}
\end{center} \vskip 1 cm

\noindent AKHLESH LAKHTAKIA$^1$\\
\noindent{$^1$Nanoengineered Materials Group, Department of Engineering Science and
Mechanics, Pennsylvania State University, University Park, Pennsylvania, USA}

\vskip 0.5 cm

\noindent TOM G. MACKAY$^2$\\
\noindent{$^2$School of Mathematics, University of Edinburgh, Edinburgh EH9 3JZ,
Scotland, United Kingdom}

\vskip 1 cm

\noindent \textbf{Abstract} \textit{Describing the phenomenon of total internal reflection
in terms of a reflection coefficient of unit magnitude, we found that,
not only can propagating plane waves be total internally reflected
at the planar interface of two
dissimilar, homogeneous, isotropic dielectric--magnetic mediums, but evanescent plane waves can also be.
The refracting medium must be the optically denser of the two mediums for total internal reflection
of an evanescent plane wave to occur.
}

\vskip 0.5  cm


Total internal reflection of a propagating plane wave at the planar
interface of two dissimilar, homogeneous, isotropic, lossless
dielectric mediums is treated in some section of virtually any
undergraduate textbook. Suppose a plane wave propagating in one
medium encounters the planar interface and the second medium is
optically rarer than the first medium; if the angle of incidence of
the plane wave is not smaller than the critical angle, total
internal reflection occurs \cite{Lorrain_Corson_1970}.

During an investigation of radiation from a dipole source embedded
in a dielectric layer \cite{Mackay_Lakhtakia_2008}, the following
question arose: Can total internal reflection of an evanescent plane
wave occur? Finding no answer to this question in the literature, we
decided to report our positive result, as well as the conditions for
an evanescent plane wave to be reflected totally.

Consider the interface of two dissimilar, homogeneous, isotropic,
lossless, dielectric--magnetic mediums,
labeled $1$ and $2$. In medium $1$ ($z\leq 0$), a plane wave with
the following electric field phasor impinges
 on the interface $z=0$:
\begin{equation}
\#E_{inc}=\left[a_s\uy +a_p\left(-\zeta_1\ux+\xi\uz\right) \right]
\exp\left[ik_1\left(\xi x +\zeta_1z\right)\right]\,,\quad z \leq 0\,.
\end{equation}
Here, $a_s$ and $a_p$ are the amplitudes
of the $s$-- and $p$--polarized
components, respectively; $k_1$ is the wavenumber in the medium;
\begin{equation}
\zeta_1=+\left(1-\xi^2\right)^{1/2}\,;
\end{equation}
and the quantity $\xi=\sin\theta$, where $\theta$ is a real--valued angle
for $\xi\in\left[0,1\right]$ and complex--valued for $\xi\in\left(1,\infty\right)$. Thus, the
plane wave is classified as either \emph{propagating} for $\xi\in\left[0,1\right]$
or \emph{evanescent} for $\xi\in\left(1,\infty\right)$.
The electric field phasor of the reflected plane wave may be set down as
\begin{equation}
\#E_{refl}=\left[r_s a_s\uy +r_p a_p\left(\zeta_1\ux+\xi\uz\right) \right]
\exp\left[ik_1\left(\xi x -\zeta_1z\right)\right]\,,\quad z \leq 0\,,
\end{equation}
where $r_s$ and $r_p$ are the reflection coefficients. We have
implicitly taken medium $1$ to support only
positive--phase--velocity propagation\footnote{For isotropic
dielectric--magnetic mediums,  the adjective `positive' (`negative')
when applied to phase velocity means that the time--averaged
Poynting vector and the wavevector are parallel (anti--parallel).
Significantly, the sign of the phase velocity determines whether
positive or negative refraction occurs \cite{MLW02}.}.

In relation to medium $1$, medium $2$ ($z\geq0$) has a relative permittivity $\epsilon_r$
and relative permeability $\mu_r$, both real--valued and both of the
same sign. If the refractive index
\begin{equation}
n_r=\epsilon_r^{1/2}\,\mu_r^{1/2}
\end{equation}
is positive, medium $2$ supports only positive--phase--velocity
propagation; if $n_r$ is negative, it supports only
negative--phase--velocity propagation. The electric field phasor of
the refracted plane wave is
\begin{equation}
\#E_{refr}=\left[t_s a_s\uy +t_p a_p\left(k_1/k_2\right)\left(-\zeta_2\ux+\xi\uz\right)\right]
\exp\left[ik_1\left(\xi x +\zeta_2z\right)\right]\,,\quad z \geq 0\,,
\end{equation}
where $t_s$ and $t_p$ are the refraction coefficients, $k_2=k_1 n_r$, and
\begin{equation}
\zeta_2=\pm \left(n_r^2-\xi^2\right)^{1/2}\,,\quad n_r\left\{\begin{array}{l}
>0\\ <0\end{array}\right.\,.
\end{equation}

After the solution of the boundary--value problem in the standard way, the  four
coefficients are obtained as follows:
\begin{equation}
\left.
\begin{array}{l}
r_s= \left({\mu_r\zeta_1-\zeta_2}\right)/\left({\mu_r\zeta_1+\zeta_2}\right)\\
r_p= \left({\epsilon_r\zeta_1-\zeta_2}\right)/\left({\epsilon_r\zeta_1+\zeta_2}\right)\\
t_s= 2\mu_r\zeta_1/\left({\mu_r\zeta_1+\zeta_2}\right)\\
t_p= 2n_r\zeta_1/\left({\epsilon_r\zeta_1+\zeta_2}\right)
\end{array}\right\}.
\end{equation}
As $r_s\equiv0$ and $r_p\equiv 0$ when $\epsilon_r=\mu_r=\pm 1$, these two cases need
not be considered.

Now,   the regime $0\leq\xi<\infty$ can be divided into five parts, depending
on whether $\zeta_1$ and $\zeta_2$ are real--valued, null--valued, or imaginary.
Then, the five
subregimes are $0\leq\xi<\xi_{a}$, $\xi=\xi_{a}$,  $\xi_a<\xi<\xi_{b}$, $\xi=\xi_{b}$, and
$\xi_b<\xi<\infty$, where $\xi_{a}={\rm min}\left\{1,\vert n_r\vert\right\}$ and
$\xi_{b}={\rm max}\left\{1,\vert n_r\vert\right\}$. In each of these subregimes, the reflection
coefficients have the following distinct characteristics:

\begin{itemize}
\item[I.] $0\leq\xi<\xi_{a}$: Both $\zeta_1$ and $\zeta_2$ are real--valued, so that
both $r_p$ and $r_s$ are real--valued.

\item[II.] $\xi=\xi_{a}$: If $\vert n_r\vert < 1$, then $\zeta_2=0\Rightarrow r_s=r_p=1$;
if $\vert n_r\vert > 1$, then $\zeta_1=0\Rightarrow r_s=r_p=-1$.

\item[III.] $\xi_a<\xi<\xi_{b}$: If $\vert n_r\vert < 1$, then $\zeta_1$ is real--valued
and $\zeta_2$ is imaginary; if $\vert n_r\vert > 1$, then $\zeta_1$
is imaginary and $\zeta_2$ is real--valued. In either case, both
$r_s$ and $r_p$ are of the form $(a-ib)/(a+ib)$ where both $a$ and
$b$ are real--valued, so that $\vert r_s\vert =\vert r_p\vert =1$.

\item[IV.] $\xi=\xi_{b}$:  If $\vert n_r\vert < 1$, then $\zeta_1=0\Rightarrow r_s=r_p=-1$;
if $\vert n_r\vert > 1$, then $\zeta_2=0\Rightarrow r_s=r_p=1$.

\item[V.] $\xi_b<\xi<\infty$: Both $\zeta_1$ and $\zeta_2$ are imaginary, so that
both $r_p$ and $r_s$ are real--valued.

\end{itemize}

Total internal reflection---interpreted as a reflection coefficient of unit magnitude---thus
occurs in
subregimes II--IV.  When $\vert n_r\vert < 1$, the refracting medium is the optically
rarer of the two mediums, and \emph{propagating} plane waves are totally
reflected provided $\vert n_r\vert\leq \xi\leq1$.  When $\vert n_r\vert >1$, the refracting medium is the optically
denser of the two mediums, and \emph{evanescent} plane waves are totally
reflected provided $1\leq \xi\leq\vert n_r\vert$.

Total internal reflection of evanescent plane
waves is exploited to frustrate the total internal reflection of
propagating plane waves and thereby create the phenomenon
of frustrated total reflection \cite{Bose,HNS2001}.

\end{document}